\documentclass[prb,aps,floats,twocolumn,eqsecnum,superscriptaddress]{revtex4}
   
\usepackage{graphicx,ifthen,amsmath,amssymb}
\usepackage{hyperref}
\hypersetup{
  dvips=true,
  latex2html=true,
  backref=true,
  pagebackref=true,
  hyperindex=true,
  colorlinks=true,
  pdfpagelabels=true,
  pdfpagelayout=TwoColumnRight,
  pdftitle={Spectrum and transition rates of the $XX$ chain analyzed via Bethe ansatz},
  pdfauthor={\textcopyright{} Michael Karbach and Gerhard Muller},
  pdfsubject={One-dimensional qunatum spin-1/2 systems},
  pdfkeywords={Heisenberg model and XX(Z) model}
}

\newcommand{\rmi}{{\mathrm{i}}}
\newcommand{\atanh}{{\mathrm {atanh\,}}}
\newcommand{\sgn}{{\mathrm{ sgn}}}

\newcommand{\KK}[2]{\mathcal{K}_{#1}(\{#2\})}

\begin{document}
\title{Spectrum and transition rates of the $XX$ chain analyzed via Bethe ansatz} 
\author{Daniel Biegel}
\affiliation{
  Bergische
  Universit{\"a}t Wuppertal, Fachbereich Naturwissenschaften, Physik,
  D-42097 Wuppertal, Germany \\
} 
 
\author{Michael Karbach}
\affiliation{
  Bergische
  Universit{\"a}t Wuppertal, Fachbereich Naturwissenschaften, Physik,
  D-42097 Wuppertal, Germany \\
}
\affiliation{
  Department of Physics, 
  University of Rhode Island, 
  Kingston RI 02881-0817, USA
} 

\author{Gerhard M{\"u}ller}
\affiliation{
  Department of Physics, 
  University of Rhode Island, 
  Kingston RI 02881-0817, USA
} 

\author{Klaus Wiele}
\affiliation{
  Bergische
  Universit{\"a}t Wuppertal, Fachbereich Physik,
  D-42097 Wuppertal, Germany \\
} 
\date{\today} 
\begin{abstract}
  As part of a study that investigates the dynamics of the $s=\frac{1}{2}$ $XXZ$
  model in the planar regime $|\Delta|<1$, we discuss the singular nature of the
  Bethe ansatz equations for the case $\Delta=0$ ($XX$ model). We identify the
  general structure of the Bethe ansatz solutions for the entire $XX$ spectrum,
  which include states with real and complex magnon momenta. We discuss the
  relation between the spinon or magnon quasiparticles (Bethe ansatz) and the
  lattice fermions (Jordan-Wigner representation). We present determinantal
  expressions for transition rates of spin fluctuation operators between Bethe
  wave functions and reduce them to product expressions. We apply the new
  formulas to two-spinon transition rates for chains with up to
  $N=4096$ sites.
\end{abstract}
\maketitle
%
\section{Introduction}\label{sec:intro}
%
The key to a meaningful interpretation of experimental or computational data for
the low-temperature dynamics of quantum many-body systems is a thorough
understanding of the nature of the physical vacuum and the dynamically relevant
collective excitations including their quasiparticle composition. In completely
integrable systems, the quasiparticle configurations that produce particular
collective excitations can be investigated close up. The identity of the former
is preserved by conservation laws notwithstanding their mutual
interaction.\cite{Taka99,KBi93} Integrable Hamiltonians that depend on
continuous parameters make it possible to observe how the physical vacuum
transforms gradually and, occasionally, changes abruptly across a quantum phase
transition. Along the way, the configurations of quasiparticles in the
collective excitations are subject to change as well.

Recently, we investigated the metamorphosis of the physical vacuum and the
dynamically relevant quasiparticles (magnons, spinons, $\psi$, $\psi^*$) of the 1D
$s=\frac{1}{2}$ Heisenberg antiferromagnet ($XXX$ model) by varying the
(integrability preserving) external magnetic field.\cite{KM00,KBM02} Introducing
instead a uniaxial exchange anisotropy also preserves integrability. The
parameter $\Delta$, which controls the anisotropy in the 1D $s=\frac{1}{2}$ $XXZ$
model,\cite{note1}
\begin{equation}\label{eq:HDelta}
  H \doteq \sum_{n=1}^N 
  \left\{S_n^xS_{n+1}^x+S_n^yS_{n+1}^y+ \Delta S_n^zS_{n+1}^z \right\} -
  \Delta \frac{N}{4},
\end{equation}
affects the physical vacuum and the quasiparticle configurations differently. In
both situations, the Bethe ansatz is an ideal framework for studying
quasiparticles, their transformations and their interactions.

The main focus in this paper is on a technical point of considerable importance
in the study of the $XXZ$ model, namely the identification of the general
structure of the Bethe ansatz solutions of all eigenstates for the case $\Delta=0$
($XX$ model). This will facilitate tracking all $XXZ$ Bethe ansatz solutions
across the planar regime, $|\Delta|<1$.

At $\Delta=0$ all states can be characterized as noninteracting composites of
fermions.  Understanding the relationship between the magnon, spinon, and
lattice fermion quasiparticles is important for the interpretation of the
excitation spectrum via dynamical probes as realized experimentally or
computationally. Furthermore, recent advances in calculating transition rates
via Bethe ansatz \cite{Kore82,IK85a,KMT99,BKM02a,BKM03} offer opportunities to
extend the list of exact results for dynamical properties of the $XX$ chain.

In Sec.~\ref{sec:II} we analyze the singularities of the $XXZ$ Bethe ansatz
equations for $\Delta\to0$. In Sec.~\ref{sec:versus} we discuss the mapping between the
spinon and fermion compositions of the $XX$ spectrum. The two-spinon spectrum of
the planar $XXZ$ model including the $XX$ limit is discussed in
Sec.~\ref{sec:2spinonXXZ}. In Sec.~\ref{sec:matel} transition rate formulas for
the $XXZ$ model are introduced and further processed for the $XX$ limit.
The two-spinon part of the dynamic spin structure factor
$S_{- +}(q,\omega)$ is discussed in Sec.~\ref{sec:2spdsf}.
%
\section{Bethe ansatz equations}\label{sec:II}
%
We consider Hamiltonian (\ref{eq:HDelta}) for even $N$ and periodic
boundary conditions over the range $0\leq\Delta\leq1$ of the anisotropy parameter. In the
invariant subspace with $z$-component of the total spin $S_T^z=N/2-r$, all
eigenstates are represented by $r$ interacting magnons. The magnon momenta $k_i$
are conserved in the scattering processes and are determined by the Bethe ansatz
equations:\cite{CG66}
\begin{equation}
  \label{eq:BA}
  e^{\rmi Nk_{i}} = \prod_{j\neq i}^r\left[
  -\frac{1+e^{\rmi(k_{i}+k_{j})}-2\Delta e^{\rmi k_{i}}}{1+e^{\rmi(k_{i}+k_{j})}-2\Delta e^{\rmi k_{j}}}\right],
  \qquad i=1,\ldots,r.
\end{equation}
The energy and wave number of any such $r$-magnon state depend on the magnon
momenta alone: 
\begin{equation}\label{eq:Erk}
  E =  \sum_{i=1}^r \left(\cos k_i - \Delta \right),\quad k = \sum_{i=1}^r k_i.
\end{equation}

At first glance it looks as if Eqs.~(\ref{eq:BA}) are drastically simplified in
the limit $\Delta\to 0$ with all magnon momenta restricted to solutions of $e^{\rmi
  Nk_i} = (-1)^{r-1}$. However, a closer look reveals that magnon pairs with momenta
$k_i+k_j\to\pi$ for $\Delta\to0$ are a common occurrence. This opens the door to nontrivial
real and complex solutions of Eqs.~(\ref{eq:BA}). The singular behavior is related
to level degeneracies. Such level crossings, which also occur at other values of
$\Delta$, have been traced back to a realization of the $sl_2$ loop algebra symmetry
at $\Delta=(q+q^{-1})/2\neq\pm1$ with $q^{2N}=1$ for $N\geq2$.\cite{DFM01,FM01,FM01a}

In the following, we use the anisotropy parameter
\begin{equation}\label{eq:gamma}
\gamma \doteq \arccos\Delta \quad (0\leq\gamma\leq\pi/2) 
\end{equation}
and transform the magnon momenta into the rapidities \cite{note2}
\begin{equation}\label{eq:yk}
  y_i \doteq \tan\frac{\gamma}{2}\cot\frac{k_i}{2}, 
  \quad i=1,\ldots,r.
\end{equation}
The Bethe ansatz equations (\ref{eq:BA}) thus become
\begin{equation}\label{eq:thety}
  \left(\frac{c_{2}y_{i}+\rmi}{c_{2}y_{i}-\rmi}\right)^{N} =
  \prod_{j\neq i}^{r}\frac{c_1(y_i-y_j)+\rmi(1-y_{i}y_{j})}{c_1(y_i-y_j)-\rmi(1-y_{i}y_{j})}
\end{equation}
for $ i=1,\ldots,r,$ with $c_1\doteq\cot\gamma$ and $c_2\doteq\cot(\gamma/2)$.  
Taking the logarithm yields
\begin{equation}\label{eq:BAEy}
  N\phi(c_2y_i) = 2\pi I_i +\sum_{j\neq i}^r\phi\left(c_1\,\frac{y_i-y_j}{1-y_iy_j}\right)
  \quad i=1,\ldots,r,
\end{equation}
with $\phi(x)\doteq 2\arctan(x)$.
Expressions (\ref{eq:Erk}) for energy and wave number now read
\begin{eqnarray}\label{eq:ey}
  E &=&
  -\frac{2}{c_{2}^{-1}+c_2}\sum_{i=1}^r\frac{y_{i}^{-1}-y_i}{(c_{2}y_{i})^{-1}+c_2y_i},
  \\ 
  \label{eq:kI}
  k &=& \pi r - \frac{2\pi}{N}\sum_{i=1}^rI_i.
\end{eqnarray}
The trigonometric Bethe ansatz equations (\ref{eq:BAEy}) have the advantage
that each solution is characterized by a set of (integer or half-integer)
Bethe quantum numbers $I_i$. These discriminating markers are used to count
and classify the solutions and are essential for numerical algorithms designed
to find solutions.

%
\subsection{$XX$-limit}\label{sec:xxepsilon-model}
%
Here we describe the general structure of the solutions of the Bethe ansatz
equations (\ref{eq:BAEy}) in the limit $\Delta\to 0$, implying $c_{1}\to0$ and $c_2\to1$.
There exist regular solutions and singular solutions. The latter are
characterized by the occurrence of pairs of rapidities $y_i,y_i'$ with the
property $y_iy_i'=1$.  For any such \emph{critical pair}, the argument of $\phi$ on
the right-hand side of Eqs.~(\ref{eq:BAEy}) is indeterminate and must,
therefore, be treated as a limit process. We shall see that some limiting
singular solutions are real while others are complex. An important fact is that
the simplified structure of the Bethe ansatz equations at $\Delta=0$ afford a
universal treatment of the singularities for arbitrary $N$.

In the absence of any critical pair of rapidities, Eqs.~(\ref{eq:BAEy}) decouple
and yield the solutions 
\begin{equation}\label{eq:1}
  y_{i}=\tan\left(\frac{\pi I_{i}}{N}\right)\quad \Leftrightarrow \quad k_i=\pi-\frac{2\pi}{N}I_i
\end{equation}
for $i=1,\ldots,r$, which are all real. The energy of any such regular state is 
\begin{equation}\label{eq:Eregu}
E = -\sum_{i=1}^r\frac{y_{i}^{-1}-y_i}{y_{i}^{-1}+y_i} = \sum_{i=1}^r\cos k_i.
\end{equation}
%
\subsection{Real critical pairs}\label{sec:real-solutions}
%
Now let us assume that among the set of rapidities $y_1,\ldots,y_r$, there is one
critical pair,
\begin{equation}
  \label{eq:2}
  y_{j}^{0}y_{j^{*}}^{0}=1\quad \Leftrightarrow \quad k_j+k_{j^*}=\pi~ ({\rm mod}~ 2\pi),
\end{equation}
and that it is a real pair. Substituting the ansatz
\begin{equation}
  \label{eq:3}
  y_{i} = y_{i}^{0} + c_1 \delta_{i},\quad i=1,\ldots,r,
\end{equation}
into Eqs.~(\ref{eq:BAEy}) and taking the limit $c_{1}\to0$ then yields the
non-critical and critical rapidities from successive orders in a $c_1$-expansion.
The non-critical rapidities $y_i$, $i\neq j,j^*$ are given by Eqs.~(\ref{eq:1}) as in
regular solutions. The $c_1$-expansion of Eqs.~(\ref{eq:BAEy}) leads to the
following equation determining the critical rapidities $y_j^0, y_{j*}^0$:
\begin{equation}\label{eq:4}
  \phi(y_{j}^{-}) =
  \frac{2}{N}\;\phi\left(\left[\frac{\sigma_{j}y_{j}^{-}}{1-\xi_{j}}\right]^{\sigma_{j}}\right).
\end{equation}
Here we have introduced $y_j^\pm=\frac{1}{2}(y_j^0\pm y_{j^*}^0)$ and we use 
\begin{eqnarray}\label{eq:sigmaxi}
   \xi_{j}  &\doteq& 
   \frac{2}{N} \sum_{i\neq j,j^{*}}^{r} 
   \frac{y_{j}^{+}-\sin k_{i}}{({y_{j}^{+}})^{-1}-\sin k_{i}},
  \\  \hspace*{-0.5cm}\label{eq:sigmaxi2}
  \sgn\!\left(\frac{y_j^+}{\xi_j}\right) \!\!&=& \!\!-\frac{2}{N}(I_j+I_{j^*}),
  \quad
  \sigma_{j} \doteq (-1)^{I_{j}-I_{j^{*}}}.
\end{eqnarray}
The non-critical $k_i$ are from (\ref{eq:1}). Criticality implies
$(y_j^+)^2-(y_j^-)^2=1$. Note that the contributions of the critical rapidities
to the energy (\ref{eq:Eregu}) cancel out. Equation~(\ref{eq:sigmaxi2})
implies that the Bethe quantum numbers of the critical pair must satisfy 
\begin{equation}\label{eq:critbgn}
|I_j+I_{j^*}|=N/2.
\end{equation}

The generalization of this solution to the case of $s$ real critical pairs, 
\begin{equation}\label{eq:5}
  y_{j_l}^{0}y_{j^{*}_{l}}^{0}=1, \quad l=1,\ldots s,
\end{equation}
with Bethe quantum numbers satisfying $|I_{j_l}+I_{j_l^*}|=N/2$ is
straightforward. The non-critical rapidities are as in Eq.~(\ref{eq:1}) and the
critical pairs are determined from
\begin{equation}
  \label{eq:6}
 \phi(y_{j_{l}}^{-}) 
    =  \frac{2}{N}\;
    \phi\left(\left[\frac{\sigma_{j_{l}}y_{j_{l}}^{-}}{1-\xi_{j_{l}}}\right]^{\sigma_{j_{l}}}\right),\quad l=1,\ldots,s 
\end{equation}
with $\xi_{j_l}, \sigma_{j_l}$ as defined in (\ref{eq:sigmaxi}) and (\ref{eq:sigmaxi2}). 

%
\subsection{Complex critical pairs}\label{sec:complex-solutions}
%
Next we consider the case of a single complex conjugate critical pair 
\begin{equation}\label{eq:7}
  y_{1}=y_{2}^{*} = u +\rmi v
\end{equation}
among the set of rapidities $y_1,\ldots,y_r$.
Equations (\ref{eq:BAEy}), rewritten as real equations, read
\begin{eqnarray}\label{eq:8}
 N\phi(c_2y_i)  &=&2\pi I_{i}
  + \sum_{ {j=3}\atop{j \neq i} }^{r} 
  \phi\left(\frac{c_1(y_i\!-\!y_j)}{1\!-\!y_iy_j}\right) 
  \nonumber \\ && \hspace{-20mm}
  +\phi\left(\frac{2c_1[(y_i\!-\!u) (1\!-\!y_i u)\!+\!y_iv^2]}%
             {(1\!-\!y_i u)^2 \!+\! (y_iv)^2 \!-\!c_1^{2}[(y_i\!-\!u)^2\!+\!v^2]}\right)
\end{eqnarray}
for $i=3,\ldots,r$ and
\begin{equation}
  \label{eq:9}
  N \phi\left( \frac{2c_2u}{1\!-\!c_2^2(u^2\!+\!v^2)} \right) =
  2\pi (I_{1}+I_{2}) + \sum_{i=3}^{r} [2\pi I_{i}-N\phi(c_2y_i)],
\end{equation}  
\begin{eqnarray} \label{eq:10}
  N \varphi \left( \frac{2c_2v}{1\!+\!c_2^2(u^2\!+\!v^2)} \right) &=&
  \varphi\left(\frac{4vc_1(1\!-\!u^2\!-\!v^2)}{(1\!-\!u^2\!-\!v^2)^2\!+\!(2vc_1)^{2}}\right) 
  \nonumber\\ &&  \hspace{-39mm}
  + \sum_{j=3}^{r}  \varphi \left( 
    \frac{2c_1v(1\!-\!y_j^2)}{(1\!-\!y_j u)^2 \!+\! (y_jv)^2 \!+\! c_1^{2}[(y_j\!-\!u)^2\!+\!v^2]} 
  \right) 
\end{eqnarray}
with $\varphi(x)\doteq 2\atanh(x)$ for the critical pair.  The $XX$ limit is performed
along the path $u = u^{0} - c_1 u^{1}$, $v = v^{0} - c_1 v^{1}$. The only
possible complex solutions that can survive the limit $c_1\to0$ are critical
pairs. The non-critical roots are again as in~(\ref{eq:1}).  The critical pair
of rapidities is now determined by the equation
\begin{equation}\label{eq:11}
  \left(\frac{1+v^{0}}{1-v^{0}}\right)^{N} =
  \left(\frac{1-\xi_{1}+v^{0}}{1-\xi_{1}-v^{0}}\right)^{2},\quad
  u_0^2+v_0^2=1, 
\end{equation}
where
\begin{eqnarray}\label{eq:12}
  \xi_{1} &\equiv& \frac{2}{N} \sum_{i=3}^{r} 
  \frac{u^{0}-\sin k_{i}}{({u^{0}})^{-1}-\sin k_{i}}, 
  \\
  && \hspace*{-1.0cm}\sgn\left(\frac{u_{0}}{\xi_{1}}\right) =-\frac{2}{N}(I_{1}+I_{2}).
\end{eqnarray}
As is custom, \cite{Taka99} we shall replace $I_1+I_2$ by a single Bethe quantum
number $I^{(*)}\pm N/2$. We then have $I^{(*)}=0$ for all complex critical pairs.

%
\subsection{Generic case}\label{sec:bethe-ansatz-equat}
%
Suppose we have $t$ critical pairs of complex conjugate solutions,
$y_{1}=u_{1}+iv_{1}=y_{2}^{*},\ldots,y_{2t-1}=u_{t}+iv_{t}=y_{2t}^{*}$, and $s$
pairs of critical real solutions, $\mathcal{J}_{l} = \{j_{l},j_{l}^{*}\} \subset
\{2t+1,\ldots,r\}$, $l=1,\ldots,s$. Then the real solutions in the limit $c_{1} \to 0$
are of the form
\begin{equation}
  \label{eq:13}
  y_{i}^{0} = \tan\frac{\pi}{N}I_{i},\quad i \notin \{1,\ldots,2t\} \cup \bigcup_{l=1}^{s} \mathcal{J}_l,
\end{equation}
\begin{equation}
  \label{eq:14}
  \phi(y_{j_{l}}^{-}) 
   =  \frac{2}{N}
   \phi\left(\left[\sigma_{j_{l}}y_{j_{l}}^{-}/(1-\xi_{j_{l}})\right]^{\sigma_{j_{l}}}\right), \quad l=1,\ldots,s,
\end{equation}
with
\begin{equation}
  \label{eq:15}
  \xi_{j_{l}} \doteq \frac{2}{N} \left[ \sum_{{i=2t+1} \atop {i\neq j_{l},{j_{l}}^{*}}}^{r}   
 \frac{y_{j_{l}}^{+}-\sin k_{i}^{0}}{{y_{j_{l}}^{+}}^{-1}-\sin k_{i}^{0}} + 2 \sum_{i=1}^{t}
 \frac{y_{j_{l}}^{+}-(u_{i}^{0})^{-1}}{{y_{j_{l}}^{+}}^{-1}-(u_{i}^{0})^{-1}} \right].
\end{equation}
For the complex solutions we obtain
\begin{eqnarray}
  \label{eq:16}
  \left(\frac{1+v^{0}_{i}}{1-v^{0}_{i}}\right)^{N} &=&
  \left(\frac{1-\xi_{i}+v^{0}_{i}}{1-\xi_{i}-v^{0}_{i}}\right)^{2},
  \\ 
  (u_{i}^{0})^2 + (v_{i}^{0})^2 &=& 1, \quad i=1,\ldots,t 
\end{eqnarray}
with
\begin{equation}
  \label{eq:17}
  \xi_{i} \doteq \frac{2}{N} \left[ \sum_{j=2t+1}^{r} 
  \frac{u_{i}^{0}-\sin k_{j}^{0}}{{u_{i}^{0}}^{-1}-\sin k_{j}^{0}} +
  2 \sum_{{j=1} \atop {j \neq i}}^{t} \frac{u_{i}^{0} - (u_{j}^{0})^{-1}}{(u_{i}^{0})^{-1} - (u_{j}^{0})^{-1}} \right].
\end{equation}
Energy and momentum of the state are determined by the non-critical roots alone:
\begin{eqnarray}\label{eq:18}
  E &=& - \sum_{{i=2t+1} \atop {i \neq j_{1},\ldots,j_{l}^{*}}}^{r} 
  \cos \left( \frac{2\pi}{N} I_i \right), 
  \\ 
  k &=& \pi(r-s-t) - \frac{2\pi}{N}\sum_{{i=2t+1} \atop {i\neq j_{1},\ldots,j_{l}^{*}}}^{r} I_{i} 
  (\mathrm{mod\;} 2\pi).\quad
\end{eqnarray}
This prescription for handling Bethe ansatz solutions in the $XX$ limit will
guarantee that all $XXZ$ Bethe eigenstates can be traced continuously across
the point $\Delta=0$. Because of the higher symmetry at $\Delta=0$ and the associated
level degeneracies,\cite{DFM01,FM01,FM01a} the relationship between the (real
and complex) Bethe eigenstates and the (always real) Jordan-Wigner eigenstates
of the $XX$ chain is nontrivial. The two representations will now be compared
for all energy levels.
%
\section{Spinons versus fermions}\label{sec:versus}
%
The full spectrum of the $XXZ$ model can be accounted for as composites of {\em
  interacting spinons} with spin 1/2 and semionic exclusion
statistics.\cite{Hald91a} For the $XX$ case an alternative and equivalent
interpretation of the complete spectrum can be established on the basis of {\em
  noninteracting spinless lattice fermions}. How are spinon
configurations related to fermion configurations in single non-degenerate
eigenstates and in groups of degenerate eigenstates?

Consider the $2^N$-dimensional Hilbert space for even $N$ divided into subspaces
characterized by $n_+$ spinons with spin up and $n_-$ spinons with spin down.
Equivalent quantum numbers are the total number of spinons and the $z$-component
of the total spin:
\begin{equation}\label{eq:npm}
2n=n_++n_-,\qquad 2S_T^z=n_+ -n_-.
\end{equation}
The dimensionality of each such subspace is \cite{Hald91a}
\begin{eqnarray}\label{eq:W}
W(n_+,n_-) &=&\prod_\sigma\left(
  \begin{tabular}{c}
    $d_\sigma+n_\sigma-1$ \\ $n_\sigma$ 
  \end{tabular}\right),
\\
d_\sigma &=& \frac{1}{2}(N+1) -\frac{1}{2} \sum_{\sigma'}(n_{\sigma'}-\delta_{\sigma\sigma'}),
\end{eqnarray}
where $\sigma=\pm$ denotes the spinon polarization. Summing $W(n_+,2n-n_+)$ over $n$ or
$n_+$ yields
\begin{equation}\label{eq:sumW}
\sum_{n=1}^{N/2}W= \left(
  \begin{tabular}{c}
    $N$ \\ $n_+$ 
  \end{tabular}\right),  \quad  \sum_{n_+=0}^{2n} W= \left(
  \begin{tabular}{c}
    $N+1$ \\ $2n$ 
  \end{tabular}\right),
\end{equation}
respectively. The double sum yields $2^N$.  In Table~\ref{tab:W} we list the
subspace dimensionalities for the case $N=8$. 
\begin{table}[htb]
  \small
  \caption{Dimensionalities $W(n_+,n_-)$ of the invariant subspaces with
    $2n$ spinons and spin $S_T^z=n_+ -n_-$ for a chain of length $N=8$. The
    equivalent quantum numbers in the fermion representation are
    $N_f=N/2-S_T^z,~ n_c=n-|S_T^z|$.}\label{tab:W}  
\begin{center}
\begin{tabular}{r|ccccc|c}
$n_+ -n_-$ & $n=0$ & $n=1$ & $n=2$ & $n=3$ & $n=4$ & $\sum_nW$\\ \hline
8~ & -- & -- & -- & -- & 1 & 1 \\
6~ & -- & -- & -- & 7 & 1 & 8 \\
4~ & -- & -- & 15 & 12 & 1 & 28 \\
2~ & -- & 10 & 30 & 15 & 1 & 56 \\
0~ & 1 & 16 & 36 & 16 & 1 & 70 \\
$-2$~ & -- & 10 & 30 & 15 & 1 & 56 \\
$-4$~ & -- & -- & 15 & 12 & 1 & 28 \\
$-6$~ & -- & -- & -- & 7 & 1 & 8 \\
$-8$~ & -- & -- & -- & -- & 1 & 1 \\ \hline
$\sum_{n_+}W$ & 1 & 36 & 126 & 84 & 9 & 256 \\ 
\end{tabular}
\end{center}
\end{table} 

\begin{figure}[htbp]
  \centering
  \includegraphics[width=80mm]{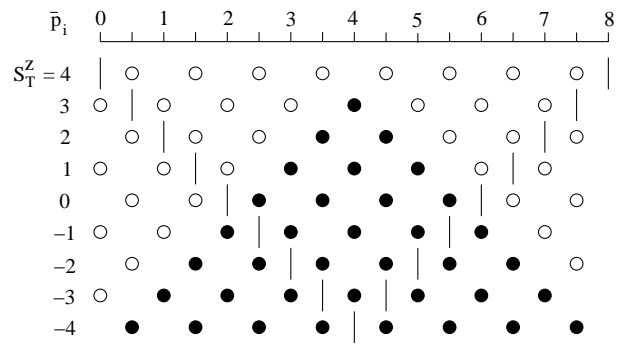}  
  \caption{Configuration in reciprocal space of the $N_f=N/2-S_T^z$ fermions in
    the lowest-energy eigenstate for given $S_T^z$ of a chain with
    $N=8$ spins. The positions of the fermions are denoted by full circles,
    those of vacancies by open circles. The fermion momenta $\bar{p}_i$ are in
    units of $2\pi/N$. The vertical bars are at wave numbers (also in units of $2\pi/N$)
    $\bar{k}_{c}^-=N/4-S_T^z/2$ and $\bar{k}_{c}^+=3N/4+S_T^z/2$.}
  \label{fig:ferm2}
\end{figure}

The $XX$ Hamiltonian in the fermion representation, transformed from
(\ref{eq:HDelta}) at $\Delta=0$ by the Jordan-Wigner mapping to a
system of free spinless fermions, reads \cite{LSM61}
\begin{equation}\label{eq:Hferm0}
H_f = \sum_p\cos p \;c_p^\dagger c_p,
\end{equation}
where the allowed values of the fermion momenta $p_i$ depend on whether the number $N_f$ of
fermions in the system is even or odd:
\begin{subequations}\label{eq:kfeo}
\begin{eqnarray}\label{eq:kfe}
p_i &\in& \{(2\pi/N)(n+1/2)\},\quad (N_f~ {\rm even}) \\ 
\label{kfo}
p_i &\in& \{(2\pi/N)n\},\quad (N_f~ {\rm odd}),
\end{eqnarray}
\end{subequations}
for $n=0,1,\ldots,N-1$.  The number of fermions $(0\leq N_f\leq N)$ is related to the
quantum number $S_T^z$ in the spin representation: $S_T^z=N/2-N_f$. No matter
whether $N_f$ is even or odd, there are $N$ distinct one-particle states.  Every
one-particle state can be either empty or singly occupied, yielding a total of
$2^N$ distinct many-particle states with energy and wave number
\begin{equation}\label{eq:Ep}
E-E_F = \sum_{i=1}^{N_f}\cos p_i, \quad k = \sum_{i=1}^{N_f}p_i.
\end{equation}
The lowest-energy state in each $S_T^z$-subspace is unique. The fermion
configuration in reciprocal space of each lowest-energy state for $N=8$ is shown
in Fig.~\ref{fig:ferm2}. 

For the further subdivision of each $S_T^z$-subspace, we introduce the wave numbers
\begin{equation}\label{eq:kcpm}
  \frac{k_c^{\pm}}{ \pi} =1\pm \left(\frac{1}{2}+\frac{S_T^z}{N}\right)
\end{equation}
shown as vertical bars in
Fig.~\ref{fig:ferm2}, dividing the band into two regions, one in the
center, one in the wings. Given that the lowest-energy state at $S_T^z\geq0$ has
all particles in the center region, we can characterize all excitations as
$n_c$-particle states. Likewise, all excitations from the lowest-energy state at
$S_T^z\leq0$ can be characterized as $n_c$-hole states. The range of this second
quantum number is $n_c=0,1,\ldots,N/2-|S_T^z|$.
The number of
fermion states characterized by $n_c$ particles or $n_c$ holes in the above sense
is then found to be expression (\ref{eq:W}) previously used for spinons, now
with 
\begin{equation}\label{eq:npmnc}
n=n_c+|S_T^z|,\quad n_{\pm}=n_c+|S_T^z|\pm S_T^z. 
\end{equation}

The next goal is to bring the mapping down to energy levels (at fixed wave
numbers) within each $(n_+,n_-)$-subspace or, in the fermion language,
$(S_T^z,n_c)$-subspace.\cite{note4} We note that for given $S_T^z\geq 0$ the number
of fermions is equal to the number of magnons in Bethe ansatz states. Moreover,
the fermion energy-momentum relation (\ref{eq:Ep}) is equivalent to the magnon
energy-momentum relation (\ref{eq:Eregu}). The exception are the critical magnon
pairs, whose momenta are different (real or complex) from the corresponding
fermion momenta (always real). Critical pairs only occur in degenerate levels
and do not contribute to the energy of the state.

This mapping between energy levels provides a useful tool for determining the
Bethe quantum numbers of eigenstates directly from the momenta of the fermion
configurations. For all non-critical rapidities, we have
\begin{equation}\label{eq:bqnf}
\frac{2\pi}{N}\,I_i=\pi-p_i.
\end{equation}
For real critical pairs, the Bethe quantum numbers are, in general, shifted
relative to the positions predicted by (\ref{eq:bqnf}), but in such a way that
$|I_j+I_{j^*}|=N/2$ is maintained. Complex critical pairs are specified by a
single Bethe quantum number $I^{(*)}=0$ as discussed in
Sec.~\ref{sec:complex-solutions}. In the following, we look more closely at the
two-spinon excitations in the two representations.

%
\section{Two-spinon spectrum}\label{sec:2spinonXXZ}
%
%
\subsection{Planar $XXZ$ model}\label{sec:2spinonXXZa}
%
Returning to the $XXZ$ model (\ref{eq:HDelta}), we note that there exist
two-spinon states with $S_T^z=0,\pm1$.  At $\Delta=1$ these states are either triplets
(with total spin $S_T=1$) or singlets ($S_T=0$). There are $\frac{1}{8}N(N+2)$
triplet levels and $\frac{1}{8}N(N-2)$ singlet levels, distinguishable by their
Bethe quantum numbers.\cite{FT81,KHM98} Integrability guarantees that each
eigenstate anywhere in the planar regime can be traced back to the point of
higher symmetry without ambiguity.  This justifies that we use the $S_T$
multiplet labels for states at $\Delta<1$.

At finite $N$ and $\Delta<1$, the two-spinon triplet components with $S_T^z=\pm1$ remain
degenerate (because of reflection symmetry in spin space) but are no longer
degenerate with the two-spinon triplet components with $S_T^z=0$. We shall see
that at $\Delta=0$, a subset of $\frac{1}{8}N(N-2)$ states from the latter set become
degenerate with the two-spinon singlets. 

We have tracked the Bethe ansatz solutions of every two-spinon state for $N=8$
from $\Delta=1$ to $\Delta=0$, indeed across this highly singular point of the Bethe
ansatz equations all the way to $\Delta=-1$.  Here we briefly report on what we found
along the stretch $0<\Delta<1$.\cite{Bieg00} For all states, solutions of the Bethe
ansatz equations were found for which all magnon momenta $k_i$ (or all
rapidities $y_i$) vary continuously across the parameter regime.

The two-spinon triplets with $S_T^z=1$ are easy to handle numerically. All
rapidities $y_i$ are finite and real.  Two-spinon triplets with $S_T^z=0$, by
contrast, tend to pose some computational challenges. At $\Delta=1$ all such states
start out with one $k_i=0$ magnon, implying $y_i=\pm\infty$. While this singular
behavior is benign,\cite{KHM98} numerical problems are caused by the fact that
in some states $y_i$ stays infinite at $\Delta<1$, whereas in other states it becomes
finite. The majority of states from this set have one real critical pair at
$\Delta=0$. To ensure continuity of the Bethe ansatz solutions when $\Delta$ is varied, it
is sometimes necessary to change the values of two Bethe quantum numbers leaving
their sum invariant.

The two-spinon singlets are far more difficult to handle numerically. Each such
state has one complex-conjugate pair of rapidities. A numerical analysis of
two-spinon singlets at $\Delta=1$ was reported in Ref.~\onlinecite{KHM98}. Again,
continuity of the solutions may make it necessary to change some Bethe quantum
numbers along the way. At $\Delta=0$ each state from this set with one
complex-conjugate critical pair becomes degenerate with a two-spinon triplet
state that has one real critical pair.

%
\subsection{$XX$ limit}\label{sec:2spinonXX}
%

\begin{figure}[htb]
  \centering
  \includegraphics[width=80mm]{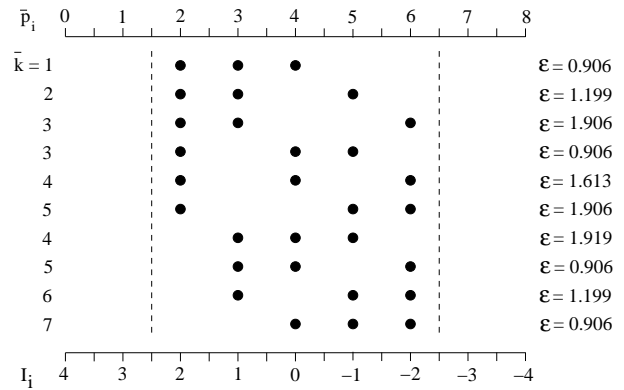}  
  \caption{Set of $N(N+2)/8$ two-spinon states with $S_T^z=1$ at $\Delta=0$ for $N=8$ as specified
    by the fermion momenta (top scale) or Bethe quantum numbers (bottom scale).
    The wave numbers $\bar{k}$ and fermion momenta $\bar{p}_i$ are in units of
    $2\pi/N$. The dashed lines represent $\bar{k}_{c}^-$ and $\bar{k}_{c}^+$. The
  energies are $\epsilon=E-E_G$ relative to the ground state with $E_G=-2.613$.}
  \label{fig:ferm4}
\end{figure}

According to the scheme developed in Sec.~\ref{sec:versus}, the two-spinon
states with $S_T^z=1$ comprise all configurations in which no state with
positive energy in the fermion band is occupied. These are the fermion
configurations with $n_c=0$. The $N/2+1$ single-particle states which are
accessible under this restriction and among which the $N/2-1$ fermions can be
distributed does indeed produce $N(N+2)/8$ distinct eigenstates. All possible
configurations for $N=8$, generated systematically across the allowed range in
reciprocal space, are depicted in Fig.~\ref{fig:ferm4}. 

The magnon momenta, rapidities, Bethe quantum numbers, and energies of these
states, now sorted according to wave numbers, are listed in Table \ref{tab:I}.
No critical pairs can be formed in any of these states, which makes it
straightforward to calculate transition rates for fairly long chains (see
Sec.~\ref{sec:2spdsf}).
\begin{table}[htb]\small
\caption{Wave number, energy, magnon momenta, Bethe quantum numbers, and
  rapidities of the $N(N+2)/8$ two-spinon triplet components with $S_T^z=1$ at $\Delta=0$ for
  $N=8$. The quantities $\bar{k},\bar{k}_i$ are in units of $2\pi/N$. 
The ground-state energy is $E_G=-2.613$.}  
\label{tab:I} 
\begin{center}
  \begin{tabular}{cc|ccc|ccc|ccc}
    $\bar{k}$ & $E-E_G$ & $\bar{k}_{1}$ &$\bar{k}_{2}$ & $\bar{k}_{3}$ 
    & $I_1$ & $I_2$ & $I_3$ & $y_1$ & $y_2$ & $y_3$ \vspace*{0.1cm} \\ \hline
1 & 0.906 & 2 & 3 & 4 & 2 & 1 & 0 & 1 & $0.414$ & 0     \\
2 & 1.199 & 2 & 3 & 5 & 2 & 1 & $-1$ & 1 & $0.414$ & $-0.414$ \\
3 & 0.906 & 2 & 4 & 5 & 2 & 0 & $-1$ & 1 & 0 & $-0.414$ \\
3 & 1.906 & 2 & 3 & 6 & 2 & 1 & $-2$ & 1 & $0.414$ & $-1$ \\
4 & 0.199 & 3 & 4 & 5 & 1 & 0 & $-1$ & $0.414$ & 0 & $-0.414$ \\
4 & 1.613 & 2 & 4 & 6 & 2 & 0 & $-2$ & 1 & 0 & $-1$ \\
5 & 0.906 & 3 & 4 & 6 & 1 & 0 & $-2$ & $0.414$ & 0 & $-1$ \\
5 & 1.906 & 2 & 5 & 6 & 2 & $-1$ & $-2$ & 1 & $-0.414$ & $-1$ \\
6 & 1.199 & 3 & 5 & 6 & 1 & $-1$ & $-2$ & $0.414$ & $-0.414$ & $-1$ \\
7 & 0.906 & 4 & 5 & 6 & 0 & $-1$ & $-2$ & 0 & $-0.414$ & $-1$\\ 
\end{tabular}
\end{center}
\end{table} 

\begin{figure}[b!]
  \centering
  \includegraphics[width=80mm]{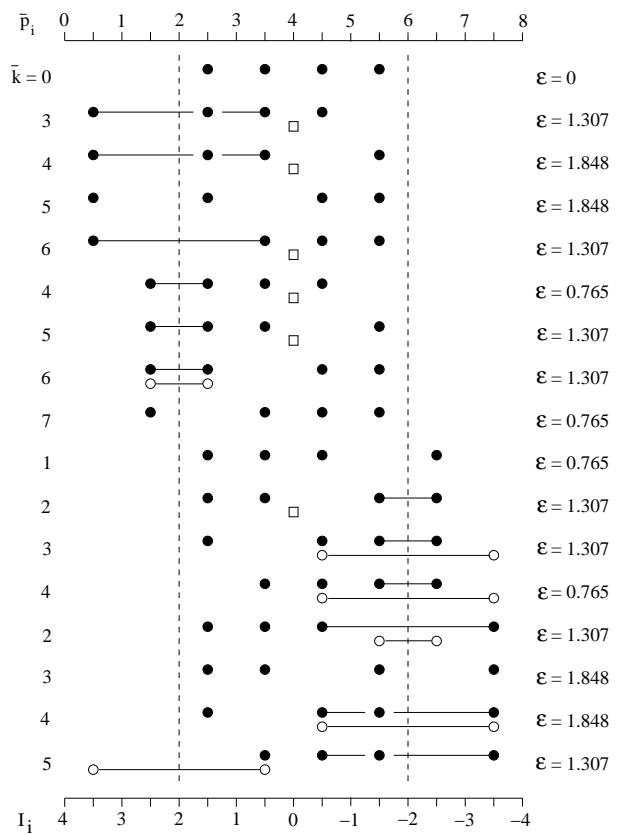}  
  \caption{Ground state and set of $N^2/4= N(N+2)/8+ N(N-2)/8$ two-spinon states
    with $S_T^z=0$ for $N=8$. The positions of the full circles in relation to
    the top scale denote the fermion momenta and those in relation to the bottom
    scale the Bethe quantum numbers predicted via Eq.~(\ref{eq:bqnf}). The
    critical pairs among them are connected by a solid line. The open circles
    and squares denote the actual Bethe quantum numbers associated with the
    critical pairs for real and complex rapidities, respectively.  The wave
    numbers $\bar{k}$ and fermion momenta $\bar{p}_i$ are in units of $2\pi/N$.
    The dashed lines represent $\bar{k}_{c}^-$ and $\bar{k}_{c}^+$.}
  \label{fig:ferm3}
\end{figure}

The two-spinon states with $S_T^z=0$ comprise all ($n_c=1$)-states 
relative to the ground state configuration in the fermion band. Removing one of
$N/2$ fermions from the ground-state configuration and placing it into one of
the $N/2$ empty single-particle states yields $N^2/4$ distinct eigenstates but
only $N(N+2)/8$ distinct levels in the $(k,E)$-plane because of degeneracies.
All possible configurations for $N=8$ including the ground state, generated
systematically across the allowed range in reciprocal space, are depicted as
rows of full circles in Fig.~\ref{fig:ferm3}. 

There are $N(N-2)/8$ twofold degenerate levels in the $(k,E)$-plane with the two
states distinguished by one pair of fermions. The two fermions in question (full
circles connected by a line in Fig.~\ref{fig:ferm3}) add up to $\pi$ mod$(2\pi)$,
thus contributing nothing to the energy (\ref{eq:Ep}). Additionally, there exist $N/2$
non-degenerate $(n_c=1)$-states at wave-numbers $Nk/2\pi=1,3,\ldots,N-1$ representing
the highest-energy two-spinon state for those wave numbers.

In summary, the $N^2/4$ $(n_c=1)$-states in the fermion representation at $\Delta=0$
originate from the set of $N(N+2)/8$ two-spinon triplets with $S_T^z=0$ and the
set of $N(N-2)/8$ two-spinon singlets. Each singlet becomes degenerate with a
triplet at $\Delta=0$, while $N/2$ triplets remain non-degenerate.  When analyzed via
Bethe ansatz, the two states of each degenerate two-spinon level with $S_T^z=0$
have one critical pair of magnon momenta as described in Sec.~\ref{sec:II}. Our
analysis shows that the critical pair is real for the triplet state and complex
for the singlet state. This is consistent with the known 2-string nature of the
two-spinon singlets and 1-string nature of the two-spinon triplets at $\Delta=1$.

In the states with real critical pairs, a solution may only exist for Bethe
quantum numbers that are shifted relative to those predicted by (\ref{eq:bqnf})
but still satisfy (\ref{eq:critbgn}). The latter are indicated in
Fig.~\ref{fig:ferm3} by open circles. In states with complex conjugate critical
pairs, two of the Bethe quantum numbers (\ref{eq:bqnf}) are replaced by a single
number $I^{(*)}=0$ representing the critical pair.

The actual Bethe ansatz solutions for $N=8$ pertaining to all $N^2/4$ two-spinon
states with $S_T^z=0$ are listed in Table~\ref{tab:II} (triplets) and
Table~\ref{tab:III} (singlets). In some instances, different configurations of
Bethe quantum numbers lead to equivalent Bethe wave functions. Only one
configuration is represented in Fig.~\ref{fig:ferm3}. \cite{note5}

\begin{widetext}
\begin{table*}[htb]
\begin{center}
\caption{Wave number, energy, magnon momenta, Bethe quantum numbers, and
  rapidities of the $N(N+2)/8$ two-spinon triplet components with $S_T^z=0$ at $\Delta=0$ for
  $N=8$. The quantities $\bar{k},\bar{k}_i$ are in units of $2\pi/N$.}\label{tab:II}  
\begin{tabular}{cc|cccc|cccc|cccc}
$\bar{k}$&$E-E_G$&$\bar{k}{_{1}}$&$\bar{k}_{2}$&$\bar{k}_{3}$&$\bar{k}_{4}$&
$I_1$ & $I_2$ & $I_3$ & $I_4$ & $y_1$ & $y_2$ & $y_3$ & $y_4$
\vspace*{0.1cm} \\ \hline
0 & 0.000 & 2.5 & 3.5 & 4.5 & 5.5 & 1.5 & 0.5 & -0.5 & -1.5 &
0.668 & 0.199 & -0.199 & -0.668 \vspace*{0.1cm} \\ \hline
1 & $0.765$ & 2.5 & 3.5 & 4.5 & 6.5 & 1.5 & 0.5 & -0.5 & -2.5 &
0.668 & 0.199 & -0.199 & -1.497 \vspace*{0.1cm} \\ \hline
2 & $1.307$ & 2.5 & 3.5 & 5.110 & 6.890 & 1.5 & 0.5 & -1.5 & -2.5 &
0.668 & 0.199 & -0.466 & -2.147 \vspace*{0.1cm} \\ \hline
3 & $1.307$ & 2.5 & 4.5 & 4.728 & 7.272 & 1.5 & -0.5 & -0.5 & -3.5 & 
0.668 & -0.199 & -0.294 & -3.402 \\
& $1.848$ & 2.5 & 3.5 & 5.5 & 7.5 & 1.5 & 0.5 & -1.5 & -3.5 & 
0.668 & 0.199 & -0.668 & -5.027 \vspace*{0.1cm} \\ \hline
4 & $0.765$ & 3.5 & 4 & 4.5 & 0 & 0.5 & -0.5 & -0.5 & -3.5 &
0.199 & 0 & -0.199 & -$\infty$ \\
& $1.848$ & 2.5 & 4 & 5.5 & 0 & 1.5 & -0.5 & -1.5 & -3.5 & 
0.668 & 0 & -0.668 & -$\infty$
 \vspace*{0.1cm} \\ \hline
5 & $1.307$ & 0.728 & 3.272 & 3.5 & 5.5 & 3.5 & 0.5 & 0.5 & -1.5 &
3.402 & 0.294 & 0.199 & -0.668 \\
& $1.848$ & 0.5 & 2.5 & 4.5 & 5.5 & 3.5 & 1.5 & -0.5 & -1.5 & 
5.027 & 0.668 & -0.199 & -0.668 \vspace*{0.1cm} \\ \hline
6 & $1.307$ & 1.110 & 2.890 & 4.5 & 5.5 & 2.5 & 1.5 & -0.5 & -1.5 &
2.147 & 0.466 & -0.199 & -0.668 \vspace*{0.1cm} \\ \hline
7 & $0.765$ & 1.5 & 3.5 & 4.5 & 5.5 & 2.5 & 0.5 & -0.5 & -1.5 & 
1.497 & 0.199 & -0.199 & -0.668 
\end{tabular}
\end{center}
\end{table*} 
\begin{table*}[htb]
\caption{Wave number, energy, magnon momenta, Bethe quantum numbers, and
  rapidities of the $N(N-2)/8$ two-spinon singlets at $\Delta=0$ for
  $N=8$. The quantities $\bar{k},\bar{k}_i$ are in units of $2\pi/N$.}\label{tab:III}  
\begin{center}
\begin{tabular}{cc|ccc|ccc|cccc}\small
$\bar{k}$ & $E-E_G$ & $\bar{k}^{*}$ & $\bar{k}_{3}$ & $\bar{k}_{4}$ & 
$I^{(*)}$ & $I_3$ & $I_4$ & $u$ & $v$ & $y_3$ & $y_4$
\vspace*{0.1cm} \\ \hline
2 & $1.307$ & 4 & 3.5 & 2.5 & 0 & 0.5 & 1.5 &
-0.765 & -0.644 & 0.199 & 0.668 \\ \hline
3 & $1.307$ & 4 & 4.5 & 2.5 & 0 & -0.5 & 1.5 &
-0.541 & -0.841 & -0.199 & 0.668 \vspace*{0.1cm} \\ \hline
4 & $0.765$ & 4 & 4.5 & 3.5 & 0 & -0.5 & 0.5 &
0 & 1 & -0.199 & 0.199 \\
& $1.848$ & 4 & 5.5 & 2.5 & 0 & -1.5 & 1.5 & 
0 & 1 & -0.668 & 0.668 \vspace*{0.1cm} \\ \hline
5 & $1.307$ & 4 & 5.5 & 3.5 & 0 & -1.5 & 0.5 & 
0.541 & 0.841 & -0.668 & 0.199 \vspace*{0.1cm} \\ \hline
6 & $1.307$ & 4 & 5.5 & 4.5 & 0 & -1.5 & -0.5 & 
0.765 & 0.644 & -0.668 & -0.199
\end{tabular}
\end{center}
\end{table*} 

\end{widetext}

%
\section{Matrix elements via Bethe ansatz for $\Delta=0$}\label{sec:matel}
%
We start from the determinantal expressions for the transition rates
\begin{equation}
  \label{eq:trarat}
  M_{\lambda}^{\mu}(q) \doteq \frac{|\langle\psi_0|S^{\mu}_q|\psi_\lambda\rangle|^2}{\|\psi_0\|^{2}\|\psi_\lambda\|^{2}},\quad \mu=z,+,-
\end{equation} 
between $XXZ$ eigenstates characterized by real $k_i$ for the spin operators
\begin{equation}\label{eq:spiflu}
S_q^\mu = \frac{1}{\sqrt{N}}\sum_n\,e^{\rmi qn}S_n^\mu,\quad \mu=z,+,-.
\end{equation}
These expressions, which were derived in Ref.~\onlinecite{BKM03} based on
previous work reported in Refs.~\onlinecite{KMT99,IK85a,Kore82} have the form
\begin{equation}\label{eq:19}
  M_{\lambda}^{z}(q) =
  \frac{N}{4} \frac{\KK{r}{y_{i}^{0}}}{\KK{r}{y_{i}}} 
  \frac{ |\det\left(\mathsf \Gamma - \frac{2}{N}\mathsf 1\right)|^{2}}%
  {\det  \mathsf K(\{y_{i}^{0}\}) \det  \mathsf K(\{y_{i}\})},
\end{equation}
\begin{equation}\label{eq:20}
  M_{\lambda}^{\pm}(q) =
  \left(
    \frac{\KK{r\pm 1}{y_{i}}}{\KK{r}{y_{i}^{0}}}
  \right)^{\pm1}\!\!\!
   \frac{|\det \mathsf{\Gamma^{\pm}}|^{2}}{\det \mathsf{K}(\{y_{i}\})\det  \mathsf{K}(\{y_{i}^{0}\})},
\end{equation}
where
\begin{equation}
  \label{eq:21}
  \mathsf{K}_{ab} = 
  \begin{cases}
      {\displaystyle \frac{\cos\gamma}{N}\frac{K(y_{a},y_{b})}{\kappa(y_{a})}}   &: a\neq b \\ 
      {\displaystyle 1-\frac{\cos\gamma}{N}\sum\limits_{j\neq a}^{r}\frac{K(y_{a},y_{j})}{\kappa(y_{a})}} &: a=b
  \end{cases}
\end{equation}
\begin{eqnarray} \nonumber
  \mathsf{\Gamma}_{ab} &\doteq & 
     F_{N}(y_{a}^{0},y_{b})
     \left(  
       \frac{1}{ G(y_{a}^{0},y_{b})}
       \prod\limits_{j=1}^{r}\frac{G(y_{j}^{0},y_{b})}{G(y_{j},y_{b})} \right. \\ \label{eq:22} 
   &&  \left.  +
       \frac{1}{ G^{*}(y_{a}^{0},y_{b})}
       \prod\limits_{j=1}^{r}\frac{ G^{*}(y_{j}^{0},y_{b})}{ G^{*}(y_{j},y_{b})}
    \right),
  \end{eqnarray}
\begin{eqnarray} \nonumber
  \mathsf{\Gamma}^+_{ab} &\doteq&   F_{N}(y_{a},y^0_b)
  \left(
    \frac{G(y_{r+1},y_{b}^{0})}{ G(y_{a},y^{0}_{b})}
    \prod\limits_{j=1}^{r} \frac{G(y_j,y^0_b)}{ G(y_j^{0} ,y_b^{0})} \right. \\ \label{eq:23} 
  && \left.  +
    \frac{G^{*}(y_{r+1},y_{b}^{0})}{G^{*}(y_{a},y^{0}_{b})}
    \prod\limits_{j=1}^{r} \frac{G^{*}(y_j ,y^0_b)}{G^{*}(y_j^{0} ,y_b^{0})} 
   \right), 
\end{eqnarray}
\begin{equation} \nonumber
  \mathsf{\Gamma}^+_{a,r+1} \doteq  1, 
  \quad a=1,\ldots r+1, \; b=1,\ldots,r,
\end{equation}
\begin{eqnarray} \nonumber
\mathsf{\Gamma}^-_{ab} &\doteq &   F_{N}(y_{a}^{0},y_b)
  \left(
    \frac{ G(y_{r}^{0},y_{b})}{ G(y_{a}^{0},y_{b})}
    \prod\limits_{j=1}^{r-1} \frac{ G(y_j^{0},y_b)}{ G(y_j ,y_b)} \right. \\ \label{eq:24} 
    && \left. +
    \frac{ G^{*}(y_{r}^{0},y_{b})}{ G^{*}(y_{a}^{0},y_{b})}
    \prod\limits_{j=1}^{r-1} \frac{ G^{*}(y_j^{0},y_b)}{ G^{*}(y_j ,y_b)} 
  \right),\\ \nonumber
\mathsf{\Gamma}^-_{ar} &\doteq&  1, 
  \quad a=1,\ldots r, \; b=1,\ldots,r-1.
\end{eqnarray}
\begin{equation} \label{eq:25}
  \KK{r}{y_{i}} \doteq  \prod_{i<j}^{r}|K(y_{i},y_{j})|,
\end{equation}
\begin{equation}  \label{eq:26}
  K(y,y') \doteq  \frac{(1-y^{2})(1-y'^{2})\sin^{2}\gamma}%
                             {(y-y')^{2} + (1-y^{2})(1-y'^{2})\sin^{2}\gamma},
                           \end{equation}
\begin{equation} \label{eq:27}
  \kappa(y) \doteq  \frac{(y^{-1}-y)\sin\gamma}{y\cot(\gamma/2) +(y\cot(\gamma/2))^{-1}},
\end{equation}
\begin{equation} \label{eq:28}
  {G}(y,y') \doteq  \frac{(y-y')\cot\gamma + \rmi (1-yy')}{\sqrt{1-y^{2}}\sqrt{1-y'^{2}}},
\end{equation}
\begin{equation} \label{eq:29}
   F_{N}(y,y') \doteq   \frac{1}{2N} 
  \frac{\sqrt{1-y'^{2}}}{\sqrt{1-y^{2}}}
  \frac{1+y^{2}+(y^{2}-1)\cos\gamma}{(y-y')\sin\gamma}.
\end{equation}
Performing the $XX$ limit, $\gamma\to\pi/2$, in these expressions is straightforward as
long as no critical pairs of rapidities are present:
\begin{eqnarray}
  \label{eq:30}
  K(y,y') &=&  \frac{(1-y^{2})(1-y'^{2})}{(1-yy')^{2} },
 \\ \label{eq:31}
  \kappa(y) &=&  \frac{y^{-1}-y}{y^{-1}+y},
  \\ \label{eq:32}
  {G}(y,y') &=&  \rmi \frac{1-yy'}{\sqrt{1-y^{2}}\sqrt{1-y'^{2}}},
  \\ \label{eq:33}
   F_{N}(y,y') & =&  \frac{1}{2N} \frac{\sqrt{1-y'^{2}}}{\sqrt{1-y^{2}}}\frac{1+y^{2}}{y-y'},
  \\ \label{eq:34} 
  \det \mathsf K & = & 1.
\end{eqnarray}
Switching back from the non-critical rapidities via (\ref{eq:1}),
$y_{i}=\cot(k_{i}/2)$, to the non-critical magnon momenta $k_i$, we can bring
expression (\ref{eq:20}) into the form
\begin{equation}
  \label{eq:35}
  {M}^+_{\lambda}(q) =  
  \frac{\prod\limits_{i=1}^{r}\prod\limits_{j=1}^{r+1}\cos^{2}\frac{k_{i}^{0}+k_{j}}{2}}%
  {\prod\limits_{i<j}^{r+1}\cos^{2}\frac{k_{i}+k_{j}}{2}\prod\limits_{i<j}^{r}\cos^{2}\frac{k_{i}^{0}+k_{j}^{0}}{2}}
  |\det \mathcal{S}^{+}|^{2},
\end{equation}
\begin{equation*}
  \label{eq:36}
 \mathcal{S}^{+}_{ab} \doteq  \left\{
  \begin{array}{ll}
    {\displaystyle \frac{2/N}{\sin k_{a}-\sin k_{b}^{0}}}       &: b=1\ldots r \\ 
     1 &: b=r+1
  \end{array}\right.,
\;\; a=1,\ldots,r+1.
\end{equation*}
\begin{equation}
  \label{eq:37}
  {M}^-_{\lambda}(q) =  
  \frac{\prod\limits_{i=1}^{r}\prod\limits_{j=1}^{r-1}\cos^{2}\frac{k_{i}^{0}+k_{j}}{2}}%
  {\prod\limits_{i<j}^{r-1}\cos^{2}\frac{k_{i}+k_{j}}{2}\prod\limits_{i<j}^{r}\cos^{2}\frac{k_{i}^{0}+k_{j}^{0}}{2}}
  |\det \mathcal{S}^{-}|^{2},
\end{equation}
\begin{equation*}
  \label{eq:38}
 \mathcal{S}^{-}_{ab} \doteq  \left\{
  \begin{array}{ll}
    {\displaystyle \frac{2/N}{\sin k_{a}^{0}-\sin k_{b}}}       &: b=1\ldots r-1 \\ 
     1 &: b=r
  \end{array}\right.,
\;\; a=1,\ldots,r,
\end{equation*}
where $\mathcal{S}^{\pm}$ are Cauchy-type matrices, whose determinants can be
evaluated explicitly:
\begin{eqnarray*}
  \label{eq:39}
  \det  \mathcal{S}^{+} & \!=\! &\left(\frac{2}{N}\right)^{r}
  \frac{\prod\limits_{i<j}^{r}(\sin k_{i}^{0}-\sin k_{j}^{0})\prod\limits_{i<j}^{r+1}(\sin k_{j}-\sin k_{i})}%
         {\prod\limits_{i=1}^{r}\prod\limits_{j=1}^{r+1}(\sin k_{j}-\sin k_{i}^{0})},
  \\ \label{eq:40}
  \det  \mathcal{S}^{-} &\!=\!& \left(\frac{2}{N}\right)^{r-1}
  \frac{\prod\limits_{i<j}^{r}(\sin k_{j}^{0}-\sin k_{i}^{0})\prod\limits_{i<j}^{r-1}(\sin k_{i}-\sin k_{j})}%
         {\prod\limits_{i=1}^{r}\prod\limits_{j=1}^{r-1}(\sin k_{i}^{0}-\sin k_{j})}.
\end{eqnarray*}
This reduces the transition rates for the perpendicular spin fluctuations (between
states without critical pairs) to compact product expressions:
\begin{eqnarray}
  \label{eq:41}
  {M}^+_{\lambda}(q) & = &
  \frac{\prod\limits_{i<j}^{r+1}\sin^{2}\frac{k_{i}-k_{j}}{2}\prod\limits_{i<j}^{r}\sin^{2}\frac{k_{i}^{0}-k_{j}^{0}}{2}}%
  {\prod\limits_{i=1}^{r}N^{2}\prod\limits_{j=1}^{r+1}\sin^{2}\frac{k_{i}^{0}-k_{j}}{2}},
  \\ \label{eq:42}
  {M}^-_{\lambda}(q) & = &
  \frac{\prod\limits_{i<j}^{r-1}\sin^{2}\frac{k_{i}-k_{j}}{2}\prod\limits_{i<j}^{r}\sin^{2}\frac{k_{i}^{0}-k_{j}^{0}}{2}} %
  {\prod\limits_{j=1}^{r-1}N^{2}\prod\limits_{i=1}^{r}\sin^{2}\frac{k_{i}^{0}-k_{j}}{2}}.
\end{eqnarray}

In the corresponding reduction of the transition rate (\ref{eq:19}) for the
parallel spin fluctuations, a complication arises, caused by the possibility
that some magnon momenta of the two states might be identical. However, this
singular behavior turns out to be instrumental for the exact evaluation of
$M_{\lambda}^{z}(q)$. A nonzero result is only possible if the two sets of Bethe
quantum numbers $\{I_i^{(0)}\}$ and $\{I_i\}$ differ by no more than one element.
For all such transitions the rate is
\begin{equation}
  \label{eq:43}
  M^{z}_{\lambda}(q) = \frac{1}{N}
\end{equation}
in agreement with a well-known result derived in the fermion
representation.\cite{KHS70}

In the following application of the transition rate expressions to a $T=0$
spin dynamic structure factor of the $XX$ model we have chosen a situation where
excited states without critical rapidities are important. The calculation of
transition rates with critical pairs requires further developmental work.

%
\section{Two-spinon transition rates}\label{sec:2spdsf}
%
From recent studies in the framework of the algebraic analysis for the infinite
chain,\cite{KMB+97,BKM98} we know that the relative integrated intensity of the
two-spinon contribution to the dynamic structure factor
\begin{equation}\label{eq:dssf}
S_{- +}(q,\omega) = 2\pi\sum_\lambda M_\lambda^-(q)\delta\left(\omega-\omega_\lambda\right)
\end{equation}
probing the spin fluctuations perpendicular to the symmetry axis of the $XXZ$
model is 73\% for the Heisenberg case $(\Delta=1)$ and steadily growing toward 100\%
on approach of the Ising case ($\Delta=\infty$).

The nonzero two-spinon intensity is reflected in the reciprocal finite-$N$
scaling behavior of the transition rates $M_{- +}^{(2)}(q,\omega_\lambda)=NM_\lambda^-(q)$ and
the scaled density of states
\begin{equation}\label{eq:sdszz}
D^{(2)}(q,\omega_\lambda)=2\pi/[N(\omega_{\lambda+1}-\omega_\lambda)],
\end{equation}
which makes the product
\begin{equation}\label{eq:prodanpm}
S_{- +}^{(2)}(q,\omega)=M_{- +}^{(2)}(q,\omega)D^{(2)}(q,\omega)
\end{equation}
converge toward a piecewise smooth function in the limit $N\to\infty$, representing the
result of the infinite chain.

\begin{figure}[tb]
  \centering
  \includegraphics[width=60mm,angle=-90]{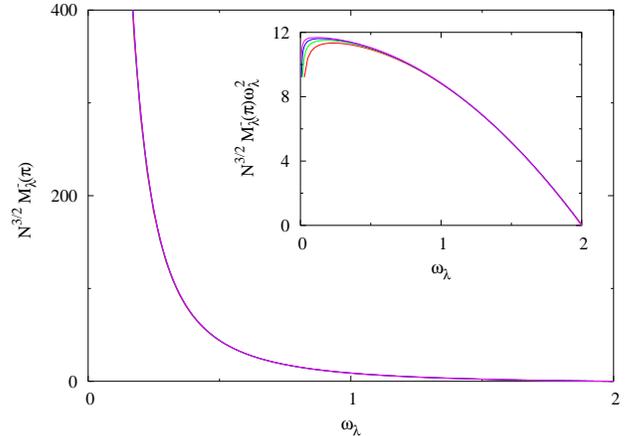}  
  \caption{Scaled two-spinon transition rates ${\bar M}_{-
      +}^{(2)}(\pi,\omega)=N^{3/2}M_\lambda^-(\pi)$ at $\Delta=0$ (data for chains of size $N=512,
    1024, 2048, 4096$). The inset shows the same quantity multiplied by
    $\omega_\lambda^2$.}
  \label{fig:mmp}
\end{figure}

A qualitatively different finite-$N$ scaling behavior is found in the $XX$ case
$(\Delta=0)$ for the two-spinon transition rates contributing to $S_{- +}(q,\omega)$. For
the transition rates to converge toward a non-vanishing piecewise smooth function
they must be scaled differently: ${\bar M}_{- +}^{(2)}(q,\omega)=N^{3/2}M_\lambda^-(q)$.
This is illustrated in Fig.~\ref{fig:mmp} for $q=\pi$. In the main plot we show
data for $N=512, 1024, 2048, 4096$ of $N^{3/2}M_\lambda^-(\pi)$ versus $\omega_\lambda$. The
scaling is near perfect across the band. The divergence building up as $N\to\infty$ in
this quantity is stronger, $\sim\omega^{-2}$, than the known infrared singularity in the
dynamic structure factor,\cite{LP75} $S_{- +}(q,\omega)\sim\omega^{-3/2}$, as documented by the inset to
Fig.~\ref{fig:mmp}. 

Given the non-reciprocal scaling behavior of the transition rates and density of
states, the relative intensity of the two-spinon dynamic structure factor $S_{-
  +}^{(2)}(q,\omega)$ vanishes in the limit $N\to\infty$. Hence the singularity structure of
the function ${\bar M}_{- +}^{(2)}(q,\omega)$ has no direct bearing on the
singularity structure of $S_{- +}(q,\omega)$. A distinct singularity structure and
spectral-weight distribution which is a property of all $2m$-spinon excitations
combined will emerge in the limit $N\to\infty$.

Consequently, the exactly known leading singularities at $\omega=0,1,2,\ldots$ in the
frequency-dependent spin autocorrelation function $\Phi_0^{- +}(\omega) =
\int_{-\pi}^{+\pi}(dq/2\pi)S_{- +}(q,\omega)$, as worked out in Ref.~\onlinecite{MS84a}, for
example, are not attributable to specific $2m$-fermion excitations, because the
integrated intensity of each $2m$-fermion contribution taken in isolation is
likely to vanish in the limit $N\to\infty$.

To reconstruct the spectral-weight distribution of $S_{- +}(q,\omega)$ at $\Delta=0$ and
to determine its singularity structure in the limit $N\to\infty$ from finite-$N$ data
for excitation energies and transition rates we need to be able to properly
handle Bethe wave functions with critical pairs of rapidities. We already know
(Sec.~\ref{sec:II}) how to solve the Bethe ansatz equations for all eigenstates
in the limit $\Delta\to\infty$. One challenging problem for the calculation of transition
rates is that Bethe wave functions with critical pairs thus obtained vanish
identically as pointed out in Ref.~\onlinecite{FM03}. However, our numerical analysis
strongly suggests that the vanishing norm $\|\psi_\lambda\|$ in the denominator of
\eqref{eq:trarat} is compensated by the vanishing transition rate in the
numerator to produce a unique finite ratio in the limit $\Delta\to0$. \cite{BKMW}


\begin{thebibliography}{29}
\bibitem[{not({\natexlab{a}})}]{note1}
\bibinfo{note}{Throughout this paper, the exchange constant is used as energy
  unit: $J=1$.}
\bibitem[{not({\natexlab{b}})}]{note2}
\bibinfo{note}{The commonly used rapidities are $z_i=2\atanh y_i$. There exist
  Bethe ansatz solutions with real magnon momenta $k_i$ for which
  $|\tanh(z_i/2)|>1$. Whereas some $z_i$ thus become imaginary, all $y_i$ stay
  real. Using the latter significantly facilitates the tasks of solving the
  Bethe ansatz equations and calculating transition rates.}
\bibitem[{not({\natexlab{c}})}]{note4}
\bibinfo{note}{A one-to-one mapping between individual states is not possible
  for degenerate levels because of the different natural eigenbases in the two
  representations.}
\bibitem[{not({\natexlab{d}})}]{note5}
\bibinfo{note}{Alternative sets of Bethe quantum numbers for the two states
  with $\bar{k}=4$ in Table~\ref{tab:III} are $(3.5,0.5,0.5,-0.5)$ and
  $(3.5,1.5,0.5,-1.5)$, respectively.}
\end{thebibliography}

\end{document}